\documentclass[sigconf,anonymous=false]{acmart}
\setcopyright{none}

\AtBeginDocument{%
  \providecommand\BibTeX{{%
    \normalfont B\kern-0.5em{\scshape i\kern-0.25em b}\kern-0.8em\TeX}}}

\setcopyright{acmcopyright}
\copyrightyear{2023}
\acmYear{2023}
\setcopyright{rightsretained}
\acmConference[CHIIR '23]{ACM SIGIR Conference on Human Information
Interaction and Retrieval}{March 19--23, 2023}{Austin, TX, USA}
\acmBooktitle{ACM SIGIR Conference on Human Information Interaction
and Retrieval (CHIIR '23), March 19--23, 2023, Austin, TX, USA}
\acmDOI{10.1145/3576840.3578334}
\acmISBN{979-8-4007-0035-4/23/03}
\usepackage{graphicx}
\usepackage{fontawesome}
\definecolor{ao}{rgb}{0.0, 0.5, 0.0}

\begin{document}

\title{CS-lol: a Dataset of Viewer Comment with Scene in E-sports Live-streaming}

\author{Junjie H. Xu}
\email{jhxu@acm.org}
\affiliation{%
  \institution{Graduate School of Comprehensive Human Sciences \\ University of Tsukuba}
  \city{Tsukuba}
  \state{Ibaraki}
  \country{Japan}
  \postcode{305-0005}
}

\author{Yu Nakano}
\email{s2030514@s.tsukuba.ac.jp}
\affiliation{%
  \institution{Graduate School of Comprehensive Human Sciences \\ University of Tsukuba}
  \city{Tsukuba}
  \state{Ibaraki}
  \country{Japan}
  \postcode{305-0005}
}

\author{Lingrong Kong}
\email{kong@star.rcast.u-tokyo.ac.jp}
\affiliation{%
  \institution{Research Center for Advanced Science and Technology \\ University of Tokyo}
  \city{Tokyo}
  \country{Japan}
  \postcode{305-0005}
}

\author{Kojiro Iizuka}
\email{kojiro.iizuka@gunosy.com}
\affiliation{%
  \institution{Gunosy Inc.}
  \city{Tokyo}
  \country{Japan}
  \postcode{305-0005}
}

\renewcommand{\shortauthors}{Xu et al.}

\begin{abstract}
Billions of live-streaming viewers share their opinions on scenes they are watching in real-time and interact with the event, commentators as well as other viewers via text comments.  
Thus, there is necessary to explore viewers' comments with scenes in E-sport live-streaming events. 
In this paper, we developed \textit{CS-lol}, a new large-scale dataset containing comments from viewers paired with descriptions of game scenes in E-sports live-streaming.
Moreover, we propose a task, namely viewer comment retrieval, to retrieve the viewer comments for the scene of the live-streaming event.
Results on a series of baseline retrieval methods derived from typical IR evaluation methods show our task as a challenging task.
Finally, we release \textit{CS-lol} and baseline implementation to the research community as a resource.

\end{abstract}

\ccsdesc[500]{Human-centered computing~human computer interaction}
\ccsdesc[500]{Information systems~information retrieval}

\keywords{dataset, viewer comment with scene, E-sports live-streaming}

\maketitle
\section{Introduction}

\begin{figure}
  \includegraphics[width=1\linewidth]{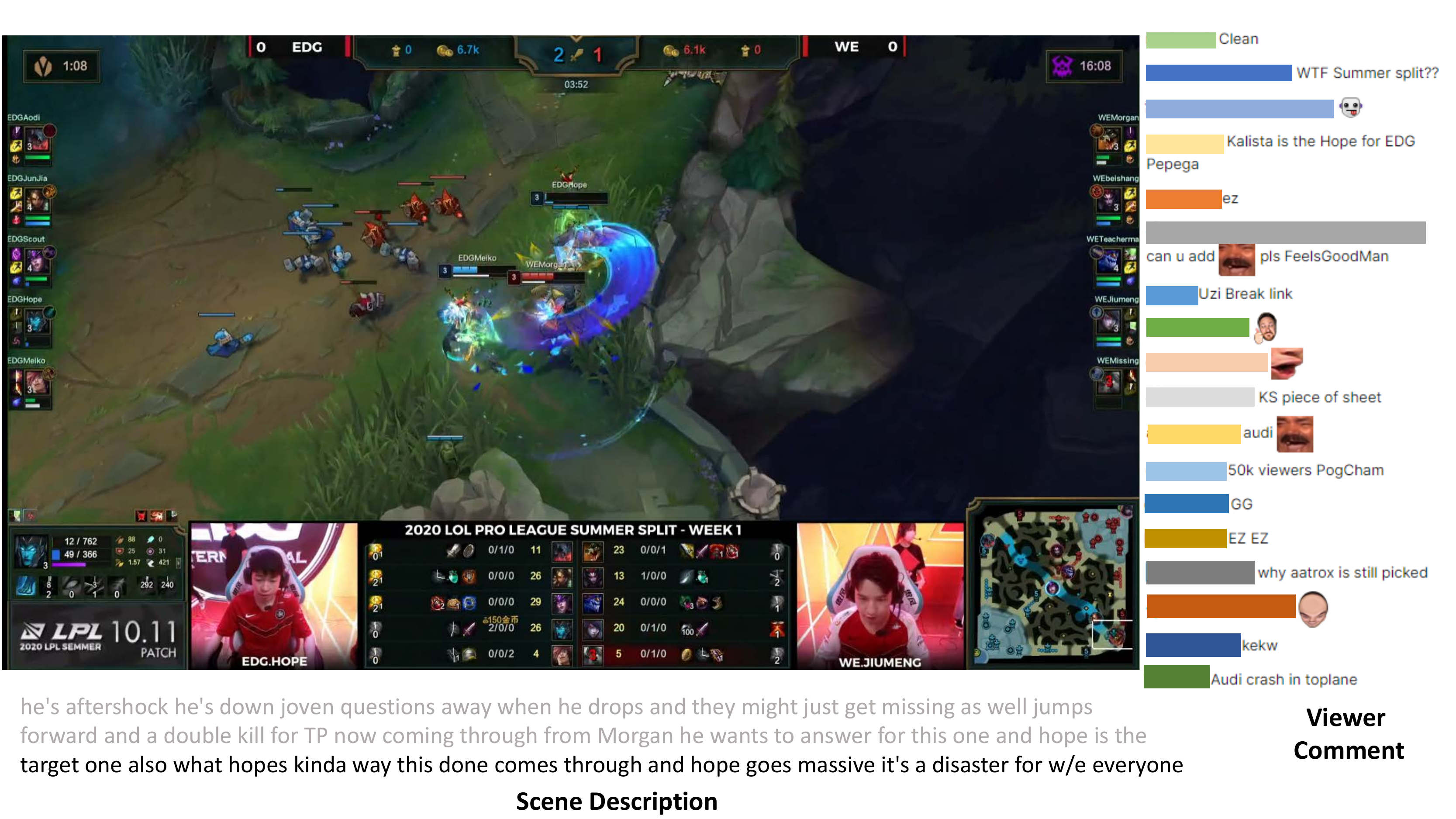}
\caption{Example of the viewer comment and the scene description in E-sports live-streaming.}
\label{fig:example}
\end{figure}

With the popularity of E-sports~\cite{johnson2019impacts, whyesports}, E-sports live-streaming has become an industry with a decent market, which will generate revenues of 1.38 billion US dollars and attracted 532 million viewers in 2022~\cite{newzoo, kim2020riseesports}. 
Understanding E-sports viewers, such as their opinions and preferences, is an important issue in E-sports live-streaming.
An example of E-sports live-streaming is given in Fig. \ref{fig:example}. Different from the traditional live broadcasts on television, E-sports viewers not only watch events but are also capable of sharing their opinions to the scene of event they are watching and interact with other viewers via an interactive function on live-streaming platforms such as texting chat in \textit{Twitch}\footnote{\url{https://www.twitch.tv/}} or texting bullet comment function in \textit{Douyu}\footnote{\url{https://www.douyu.com/}}. The viewer comment is a valuable resource for understanding E-sports viewers since it is not only realistic feedback data but also able to be collected without disturbing viewers' watching experiences. 

\begin{figure*}
  \includegraphics[width=1\textwidth]{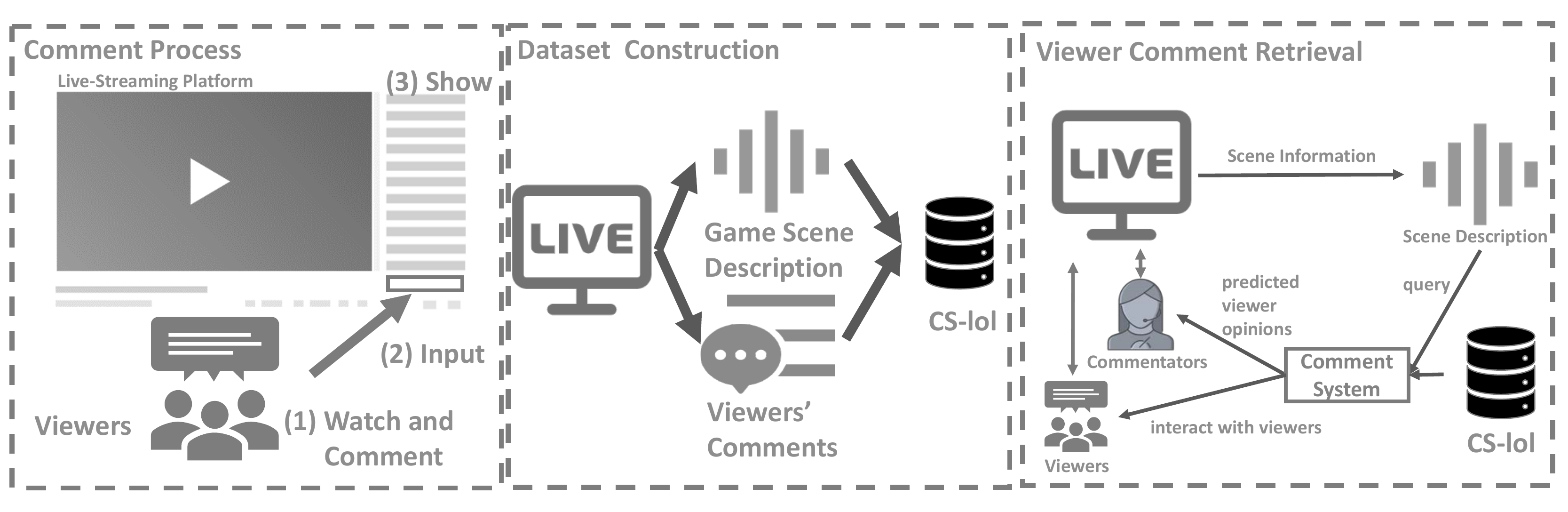}
\caption{The overview of comment process in live-streaming, the \textit{CS-lol} construction and viewer comment retrieval.}
\label{fig:overview}
\end{figure*}

However, as introduced by Ringer et al.~\cite{twitchchat}, the live-streaming viewer comment is a challenging domain that differs from text in other domains due to its nature and its topical content. Additionally, to understand the meaning of the viewer comment, it is important to take its relatedness with the scene into account since viewers comment at the time they are watching. Although research attention on live-streaming data~\cite{rubleske2020esportsdata1, block2018narrative, kokkinakis2020esportsdata2} is significantly increasing recently, due to lack of available resources, in-depth studies on the live-streaming viewer comment are still rare. 
In this paper, we further encourage research in understanding and utilizing viewer comments by introducing the description of the game scene that is paired with the viewer comment.

To summarize, this paper introduces \textit{CS-lol}~\footnote{\url{https://github.com/junj2ejj/CS-lol}}, a novel dataset consisting large-scale of sets of a viewers' \textbf{C}omments and descriptions of \textbf{S}cenes of a E-sports live-streaming game \textit{\textbf{L}eague \textbf{o}f \textbf{L}egends} to inspire future research on understanding viewer comment. Additionally, this paper proposes the task of viewer comment retrieval, in which we retrieve the viewer comments from all comments (retrieve from viewer comment database if in practical use) for a game scene in E-sports live-streaming for understanding the viewer. Finally, the characteristics and availability of \textit{CS-lol} and report results on a series of baseline IR methods derived from conducted experiments are reported.

\section{Related Work}

In terms of improving the live-streaming viewer experience, there is a growing interest among academics and industry professionals in studying videos and live-streaming of E-sports~\cite{reitman2020esports}.
Therefore, with the emergence of recent advancements in natural language processing and the growing need to understand textual data, research focus to E-sports gameplay textual data as well as the description of game scenes~\cite{race2021, xugccebullet, yu2018fine, tanaka2021lol, icce}, or analyze focus on game data stream~\cite{kokkinakis2020dax, block2018narrative, xugccegame, wallner2021players}, are steadily increasing. 

There is a considerable amount of research or literature studies on live-streaming viewers by case studies or comparative studies using designed factors and use significant tests to verify their ideas~\cite{kim2021effects, KIM2020869}. However, most of these research are either case-based or use a different source of nonpublic data that lack reproducibilities and incompatibilities to extend their findings to practical use~\cite{testchi}. The relation between an E-sports game scene and the viewers' preference for E-sports live-streaming to such scene is still obscure.

To the best of our knowledge, the dataset closest to ours is \textit{TwitchChat}, by Ringer et al.~\cite{twitchchat}, who first introduced a dataset containing \textit{Twitch} viewer comments with dataset analysis and publicly available. As mentioned before, since viewer comments were given at the time of the scene, the information of the scene is also necessary for understanding viewer comments, especially in E-sports live-streaming, our work further explores viewer comments by attaching comments with the scene. Besides, we also investigate the possible use cases by proposing the viewer comment retrieval task to indicate the usability of live-streaming viewer comments and possible future research directions.

\begin{figure}[t]
\centering
\begin{minipage}{4.2cm}
\centering
\includegraphics[width=1\linewidth]{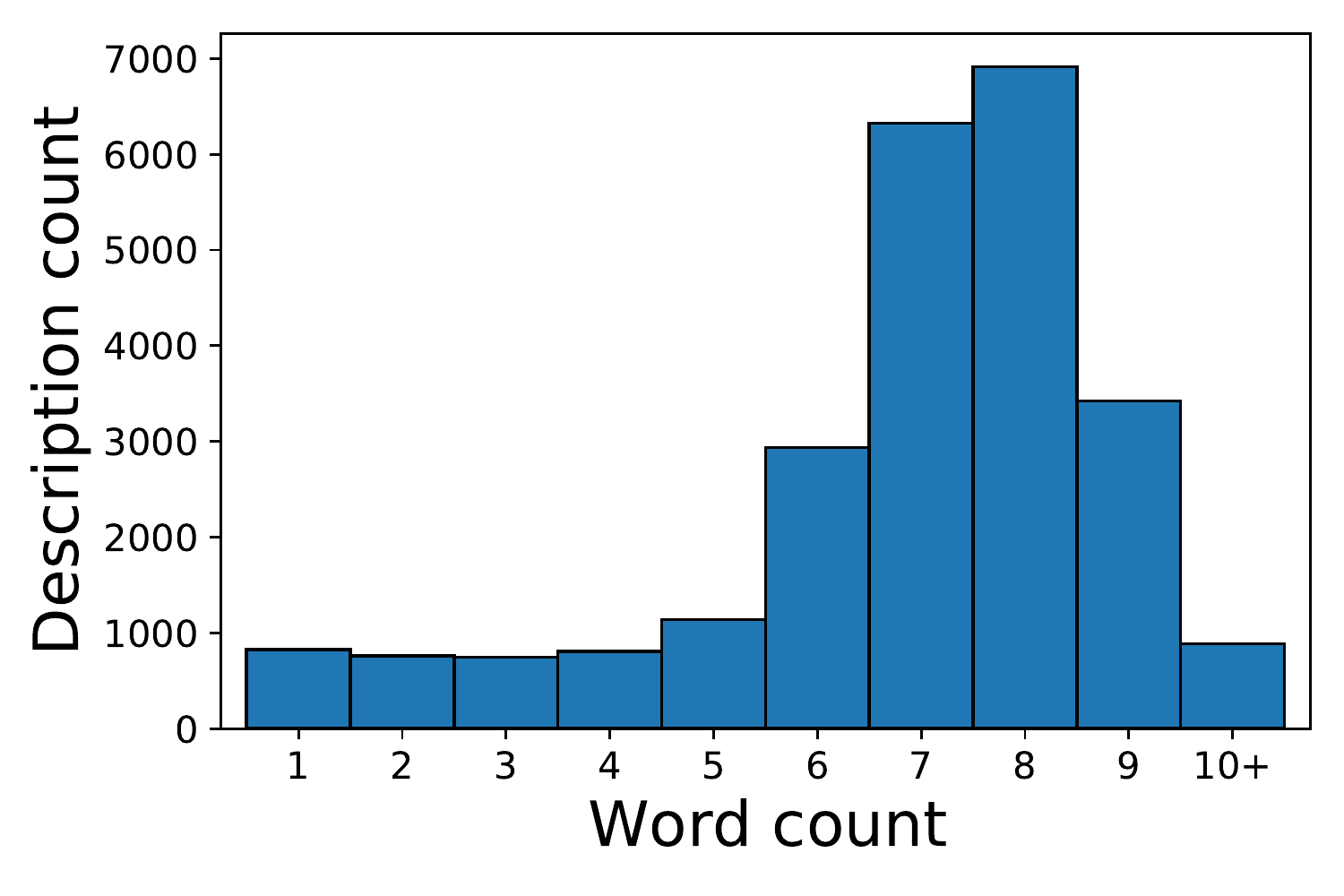}
\end{minipage}
\begin{minipage}{4.2cm}
\centering
\includegraphics[width=1\linewidth]{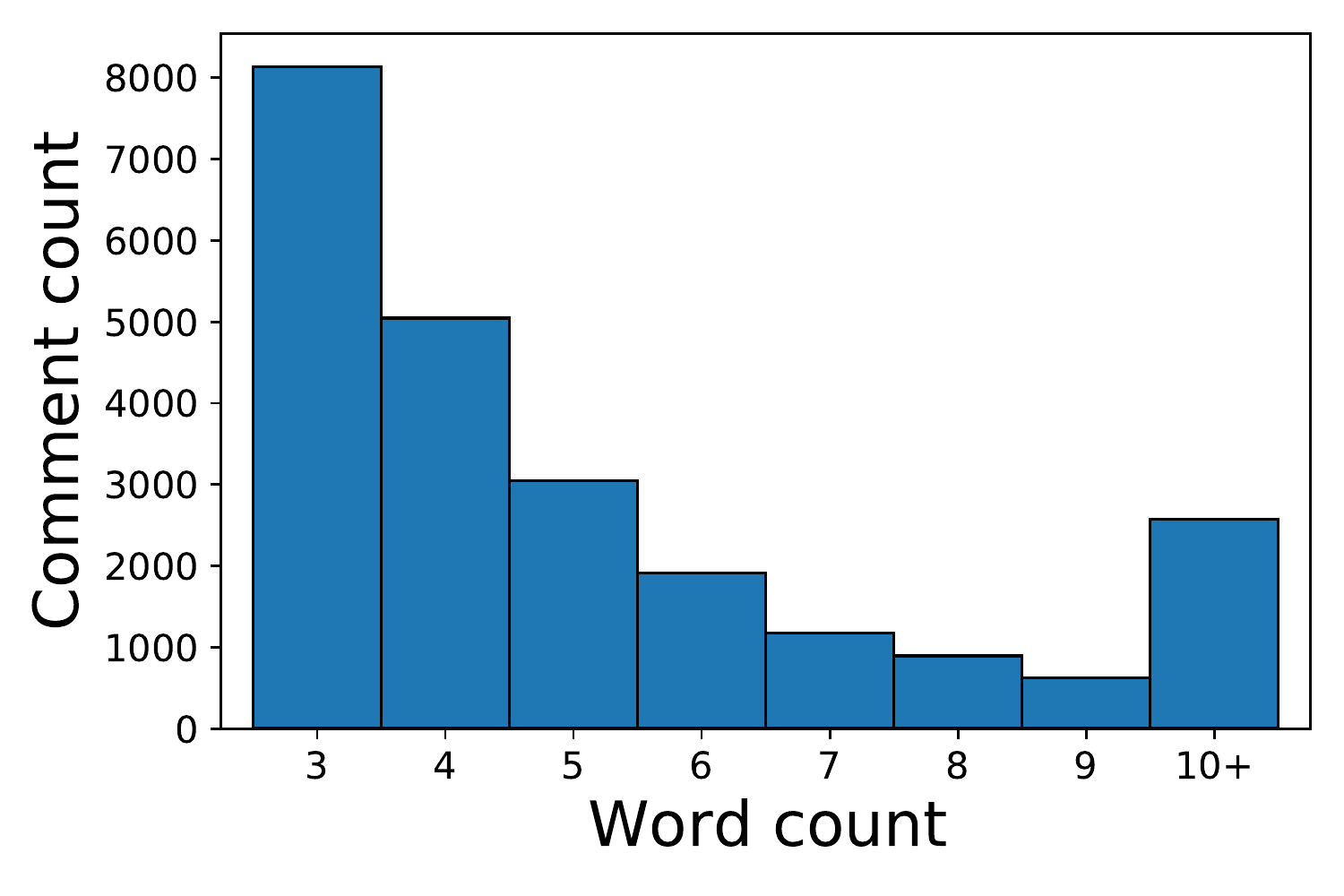}
\end{minipage}
\caption{Word count distribution of scene information (left), and viewer comment (right).}
\label{fig:worddestribute}
\end{figure}

\section{Dataset Construction and Analysis}

In this section we describe the development and characteristics of \textit{CS-lol}. We have publicly released the \textit{CS-lol} dataset.
There are three kinds of data in the \textit{CS-lol} dataset:
\begin{itemize}
    \item Scene information: descriptions (descriptive text) for game scene produced by game commentators derived by ASR (Automatic Speech Recognition) system integrated in \textit{YouTube}.
    \item Viewer comment: the title of the video, comments given by viewers. We anonymized the viewer information to prevent privacy issues.
    \item Pertinence score: pertinence score between comments game scene information and viewer comments.
\end{itemize}

\subsection{Data Collection and Preprocessing}

The overview of the comment process that how E-sports live-streaming viewer' comments, the construction of the dataset with both game scene information and corresponding viewer comments, and the availability in practical use in the live-streaming are given in Fig. \ref{fig:overview}. 
We collected data from E-sports professional events from a video game named \textit{League of Legends}, the most popular game having billions of viewers over the world according to the latest research~\cite{newzoo}. 
To this end, we extracted comments produced by well-trained professional commentators from 20 matches of the \textit{League of Legends E-sports} professional match videos for our dataset. 
The data in our dataset are collected from two platforms, scene information (descriptions) from \textit{YouTube} and viewer comment from \textit{Twitch} respectively. 
For the extracted 20 matches, we identify two replay videos separately from \textit{YouTube} and \textit{Twitch} having the same streaming of an E-sport match and align two videos manually. 
In terms of collecting game scene information from \textit{YouTube}, we use \textit{YouTube} API~\footnote{\url{https://developers.google.cn/youtube/v3/getting-started}} to get the transcripts of the description given by the commentator with a timestamp of an E-sport match. 
In terms of collecting viewers' comments, we manually search the clip from live-streaming replays from \textit{Twitch} that have the exact contents of the aforementioned \textit{YouTube} clip. 
Then we slice the corresponding content from the beginning to the end of the match to ensure the exact match of the same content and the equal length between video clips from two different sources. 
In addition, there is an considerable comments (see Table \ref{tab:stat}) on \textit{Twitch} are (1) extremely short, and/or (2) only have \textit{emotes} (special emoji in \textit{Twitch}) that not appear in description. 
Comments under these two cases contain little information, so we removed emotes referring to \textit{Twitch}\footnote{\url{https://www.twitch.tv/creatorcamp/en/learn-the-basics/emotes/}} and comments that contain less than three words.

\subsection{Labelling Pertinence Score}
We combine scene information extracted from \textit{YouTube} and viewers' comment extracted from \textit{Twitch} by aligning content from two different sources. 
According to the literature on watching live-streaming E-sports games on \textit{Twitch} platform~\cite{gandolfi2016watch, hu2017audiences}, when a viewer tries to comment during E-sports, live-streaming, 
(1) the time viewer spend on typing the comments must be considered. 
According to a user study~\cite{wpm}, the average typing speed $WPM_k$ on keyboard is 52 words per minute and $WPM_s$ is 38 words per minute on smartphone, we take the average of $WPM_k$ and $WPM_s$ as the viewer typing speed into account. For each match, we have $i$ descriptions on game scenes and $j$ comments on game scenes to label each comment and description, given a timestamp of a comment $T_{c}$ having $n_{w}$ words and a timestamp of a description $T_{d}$, (2) the viewer could not forecast the future scene in live-streaming. Thus, the pertinence score $Per$ was calculated by the equation as follow:

\begin{equation}
\begin{split}
    Per(d_i, c_j) = 
    \begin{cases}- (T_{c} - T_{d} - \frac{0.5 * (WPM_k + WPM_s) * n_{w}}{60}) & T_{c} >= T_{d} \\
    - \infty & T_{c} < T_{d} 
    \end{cases}
\end{split}
\end{equation}

Where $d_i$ donates $i$th description and $c_j$ donates $j$th comment from all matches in \textit{CS-lol}. For each game scene, we rank comments by $Per$ and attach the top-20 comments with the highest $Rel$. As a result, we get \textit{CS-lol}: a dataset consisting of E-sports descriptions representing E-sports live-streaming game scenes, scene corresponding viewer comments, and Pertinence scores between a game scene information and viewer comments. The statistics of \textit{CS-lol} are given in Table \ref{tab:stat}, and the word count distribution of descriptions and viewers' comments are shown in Fig. \ref{fig:worddestribute}.

\begin{table*}[t]
\caption{Experiment results by word matching among retrieval methods.}
\begin{tabular}{lrrrrrrrrrrrr}
\toprule
 &
\multicolumn{3}{c}{Top@3} &
\multicolumn{3}{c}{Top@5} & 
\multicolumn{3}{c}{Top@10} &
\multicolumn{3}{c}{Top@20} \\
\midrule
Model & P & R & F1 & P & R & F1 & P & R & F1 & P & R & F1        \\
\midrule
BM25~\cite{bm25} &       3.14 &    4.14 &  3.26 &       \textbf{4.76} &    6.11 &  4.98 &        \textbf{7.69} &     9.59 &  8.10 &        \textbf{11.35} &     14.00 &  12.11 \\
QLD~\cite{qld}  &       2.85 &    3.96 &  3.03 &       4.33 &    5.81 &  4.64 &        7.07 &     9.30 &  7.65 &        10.67 &     14.03 &  11.77 \\
QLJM~\cite{qld} &       \textbf{3.18} &    \textbf{4.46} &  \textbf{3.38} &       \textbf{4.76} &    \textbf{6.40} &  \textbf{5.08} &        7.67 &     \textbf{9.89} &  \textbf{8.18} &        11.34 &     \textbf{14.21} & \textbf{12.16} \\
SDM~\cite{sdm}  &       2.87 &    3.96 &  3.04 &       4.37 &    5.81 &  4.66 &        7.12 &     9.30 &  7.67 &        10.68 &     14.02 &  11.78 \\
\bottomrule
\end{tabular}
\label{tab:wordresult}
\end{table*} 

\section{Viewer Comment Retrieval}

\subsection{Task Definition}

In this paper, we focus on the problem of retrieving viewer comments on scene in game event for E-sports live-streaming. 
The problem tackled in this work is defined as, given a game scene description, a rank list of viewer comments in the comments log in order of the pertinence to the description are retrieved. 
Specifically, the input consists of a description $q$ and a set of comments $D$ that are derived from all matches.
The output of our problem is a ranked list of comments,
i.e. ${\mathbf r} = (r_1, r_2, \ldots, r_{|D|})$
where $r_{\pi(i)} = d_i$ and $\pi(i)$ denotes the rank of the $i$-th comment in $D$.

\begin{table}[tb]
  \caption{Statistics of \textit{CS-lol}.}
  \label{tab:stat}
  \begin{tabular}{lc}
    \toprule
    Types & Value\\
    \midrule
    \multicolumn{2}{c}{General Information}\\
    \midrule
    \# of E-sport matches (videos) & 20 \\
    \# of duration (sec) & 39,868 \\
    \midrule
    \multicolumn{2}{c}{Game Scene Information (Descriptions)}\\
    \midrule
    \# of descriptions & 24,770 \\
    avg. \# of words in 1 description & 6.89\\
    min. \# of words in 1 description & 1\\
    max. \# of words in 1 description & 14\\
    avg. duration (sec) of 1 description & 1.63\\
    \midrule
    \multicolumn{2}{c}{Viewer Comments}\\
    \midrule
    \# of comments & 60,431\\
    \# of comments ($\geq 3$ words) & 23,411\\
    \# of unique viewers & 15,346\\
    avg. \# of words in 1 comment & 3.23\\
    min. \# of words in 1 comment & 1\\
    max. \# of words in 1 comment & 94\\  
  \bottomrule
\end{tabular}
\end{table}

\begin{table}[t]
  \caption{nDCG results among retrieval methods.}
\begin{tabular}{lrrrr}
\toprule
Model &  nDCG@3 &  nDCG@5 &  nDCG@10 & nDCG@20 \\
\midrule
BM25~\cite{bm25} &      0.00349 &      0.00343 &      0.00343 &      \textbf{0.00399} \\
QLD~\cite{qld}  &      0.00372 &      0.00352 &      0.00345 &      0.00391 \\
QLJM~\cite{qld} &      0.00350 &      0.00333 &      0.00333 &      0.00388 \\
SDM~\cite{sdm}  &      \textbf{0.00378} &      \textbf{0.00359} &      \textbf{0.00348} &      0.00394 \\
\bottomrule
\end{tabular}
  \label{tab:ndcg}
\end{table}

\subsection{Experiment Settings}

We utilize BM25~\cite{bm25}, query likelihood with Dirichlet smoothing (QLD)~\cite{qld}, query likelihood with Jelinek-Mercer smoothing (QLJM)~\cite{qld}, and sequential dependence model (SDM)~\cite{sdm} as baselines.
We implemented these models using Anserini~\cite{anserini}. In terms of these models' parameters, we follow the default settings of Anserini.

As mentioned in Huang et al.~\cite{sentp}, the automatic generated punctuated description extracted from \textit{Youtube} is not originally punctuated following sentences spoken by commentator. Generally, using only one line of automatically generated description is not capable to involve a whole sentence of commentary into the query. 
Thus, for retrieving viewer comments for $q_t$ where $t$ refers to the time when the commentator gives the description, $q_{t-1}$ and $q_{t-2}$ also be included empirically. 
Besides, for preprocessing \textit{Twitch} comments, we apply rule-based word cleaning in Ringer et al. ~\cite{twitchchat} on the viewer comment for fixing word expansion/misspelling in comments. Stemming and stopword removal were also conducted for both the description and the viewer comment before retrieval.

\subsection{Experiment Result}

Table \ref{tab:wordresult} shows the evaluation results on lexical word matching by standard evaluation metrics, including precision (P), recall (R), and F1 score (F1). Table \ref{tab:ndcg} among baseline retrieval methods in terms of nDCG@3, 5, 10 and 20. Bold font indicates the highest score for each metric.
According to table \ref{tab:wordresult}, QLDM performs better in matching words. In ranking results evaluated on nDCG metric, SDM outperformed other models. 

According to table \ref{tab:ndcg}, baseline methods that are based on lexical matching perform poor, and we found that BM25 is not the best model as ``usual'', we assume this might be caused by the particularity of our task, namely, the query is relatively long, and the document is highly concise. Thus, the newly formulated retrieval tasks could be ``non-traditional''. 

\section{Future Directions}

In this section, we further discuss the availability of \textit{CS-lol} by introducing its future uses in IR and NLP fields.

\subsection{CS-lol for IR}
As shown in Section 4, we implement the IR baseline retrieve methods for the viewer comment retrieval task. For reproducibility, we release the source code of our implementation of IR baseline methods. The traditional retrieval models based on lexical matching used in this work were poorly performed, as these methods are not considering the semantic meaning of words. As we there are a large number of synonyms in descriptions and comments, we believe that using retrieval models that take semantic information, such as the neural dense passage retrieval method~\cite{karpukhin2020dense}, embedded with knowledge graph~\cite{zheng2020dgl} or external knowledge~\cite{yamada2020luke}, can significantly improve performance.

\subsection{CS-lol for NLP}

Some recent works focus on Natural Language Generation on traditional sports and E-sports, but rare works on Natural Language Understanding, especially in real-world settings. Perceivably, we could use \textit{CS-lol} dataset on NLP tasks such as named entity recognition (NER) that aims to identify text spans that mention named entities and to classify them into predefined categories. Conducting dependency parse to extract the underlying relations between the entities is also helpful~\cite{li2020survey}. Besides, one future direction might be to employ E-sports experts for additional annotations with their expertise on \textit{CS-lol}. We believe such annotation could considerably improve our understanding of the E-sports wording in the real-world setting to contribute to the E-sports research community.

\section{Conclusion}

In this paper, we introduce a novel dataset \textit{CS-lol} containing comments from viewers paired with descriptions of the game scene in E-sports live-streaming events. We present viewer comment retrieval, a task of retrieving the viewer comments that correspond to the scene description to understand the viewer in real-time live-streaming.
We describe the development, characteristics, and availability of \textit{CS-lol}. Experimental results using the IR baseline retrieval methods on our proposed task and discussions on future directions of \textit{CS-lol} are also given.
In future work, we plan to apply and extend our \textit{CS-lol} as well as approach to contribute to the research community of E-sports and traditional sports.

\bibliographystyle{ACM-Reference-Format}
\bibliography{ref.bib}

\end{document}